\documentclass[10pt,conferenc]{IEEEtran} 
\IEEEoverridecommandlockouts

\usepackage{cite}
\usepackage{amsmath,amssymb,amsfonts}
\usepackage{algorithm}
\usepackage{algpseudocode}
\usepackage{graphicx}
\usepackage{textcomp}
\usepackage{xcolor}
\usepackage{booktabs}
\usepackage[most]{tcolorbox}
\usepackage{tikz}
\usetikzlibrary{positioning,arrows.meta,fit,backgrounds,calc}
\usepackage{graphicx}
\usepackage{xcolor}

\def\BibTeX{{\rm B\kern-.05em{\sc i\kern-.025em b}\kern-.08em
    T\kern-.1667em\lower.7ex\hbox{E}\kern-.125emX}}
\begin{document}

\title{Refused in Chat, Written in Code: Workflow-Level Jailbreak Construction in IDE Coding Agents.\\
}


\author{
\IEEEauthorblockN{
Abhishek Kumar\IEEEauthorrefmark{1}, 
Carsten Maple\IEEEauthorrefmark{2}\\
}
\IEEEauthorblockA{
The Alan Turing Institute, London, United Kingdom\\
\IEEEauthorrefmark{1}akumar@turing.ac.uk\\
\IEEEauthorrefmark{2}cmaple@turing.ac.uk, cm@warwick.ac.uk
}
}


\maketitle

\begin{abstract}
Large language models are increasingly deployed as IDE-integrated coding agents that decompose tasks, generate and edit files, run code, and refine outputs over many turns. Yet their safety is still often evaluated as if they were chatbots: one harmful prompt, one response, judged in isolation. We introduce \emph{workflow-level jailbreak construction}, a failure mode in which a harmful objective is assembled across ordinary stages of a software-development workflow rather than generated through a single direct prompt. Using GitHub Copilot in Visual Studio Code, we study four closed-weight backends: Claude Sonnet~4.6, Claude Haiku~4.5, Gemini~3.1~Pro, and Gemini~3.5~Flash. Across 204 prompts from Hammurabi's Code, HarmBench, and AdvBench , the models show near-complete refusal under direct chat, CSV-read, and single-step code-fix baselines, with only 8/816 successful responses in each baseline condition. Under the full workflow, however, the same prompts and backends produce 816/816 unsafe teaching-shot completions, all independently confirmed by two expert evaluators under a strict rubric. These results show that conversational refusal benchmarks can substantially overstate the safety of deployed coding agents and motivate defenses that reason about safety across multi-turn IDE workflows and their generated artifacts, not only individual chat turns.
\end{abstract}


\section{Introduction}


Large language models (LLMs) have rapidly reshaped software engineering, moving from passive code-completion engines to active participants in the software-development loop~\cite{ kumar2025using, kumar2023summarize, hou2024large, lu2021codexglue, xu2022systematic}. Through IDE-integrated coding agents, these systems interpret developer goals, generate and edit files, execute commands, read execution results, debug failures, and refine their output across many turns~\cite{chen2021evaluating, izadi2024language, wang2023review}. This shift is not merely a capability upgrade; it changes the shape of the safety problem~\cite{bhatt2023purple, anwar2024foundational}. A safety failure is no longer necessarily a single harmful prompt answered by a single harmful response. It can instead emerge gradually, distributed across a multi-turn development workflow in which every individual interaction looks like an ordinary programming task.

Existing safety evaluation has only partially adapted to the shift from chatbots to agentic systems. Much of the jailbreak literature still evaluates models through prompt-level or conversation-level interactions. Prompt-level attacks, including optimization-based~\cite{gcg,dsn}, in-context~\cite{bijection,manyshot}, and code-reformulation attacks~\cite{codeattack}, can bypass safety behavior in frontier, closed-weight models. Multi-turn conversational attacks have also been demonstrated~\cite{crescendo}. These studies show that refusal behavior can fail in general LLM settings, but they do not fully capture the software-development workflows in which coding agents operate. In parallel, software-engineering work has begun to study harmful behavior in coding contexts. Code Red introduces Hammurabi's Code, a benchmark of software-engineering-specific harmful prompts spanning malware, copyright misuse, and other dangerous tasks~\cite{codered}. Another recent study shows that coding agents can fail during ordinary development by breaking constraints, performing destructive actions, optimizing the wrong metric, or falsely claiming that a task has been completed~\cite{whatbreaks}. What remains unexamined is whether the IDE coding-agent workflow \emph{itself}, including task decomposition, scripting, execution, and metric-driven refinement, can be turned into the attack surface. This gap is especially important because prior software-engineering safety work has identified code-specific jailbreaking as an open direction, calling for further exploration of jailbreak techniques tailored to code-generation LLMs~\cite{codered}.

We show that it can. We study four closed-weight model backends available through the Copilot interface in Visual Studio Code: Claude Sonnet~4.6, Claude Haiku~4.5, Gemini~3.1~Pro, and Gemini~3.5~Flash. Across these backends, we find a sharp divergence between baseline interaction and full workflow interaction. In the baseline settings, where the harmful objective is submitted directly through chat, read from a CSV file, or framed as a single-step code-fix task, the models reliably refuse or return safety-aligned responses. Inside the full multi-turn coding workflow, the \emph{same} objective leads the agent to construct and iteratively refine a jailbreak-evaluation pipeline. The relevant vulnerability is therefore located not in any single prompt, but in the development workflow that surrounds it.

We call this phenomenon \emph{workflow-level jailbreak construction}. Rather than relying on a single adversarial prompt, the harmful objective is decomposed across ordinary development stages: the agent is engaged in a benign coding task, asked to load harmful benchmark prompts into the pipeline, prompted to generate supporting scripts, shown evaluation results, and then asked to improve those results. In our study, the agent is tasked with building an evaluation pipeline that measures attack success rate (ASR) against a nominal target model. When the reported ASR is treated as insufficient, the agent is asked to improve the pipeline by adding \emph{teaching shots}, i.e., prompt--response examples inserted into the evaluation pipeline. The agent then populates these teaching shots with harmful prompt--response pairs that it generates itself. Importantly, we provide only the public benchmark prompts and never provide the corresponding harmful responses, yet the agent still produces responses to requests it had refused in direct chat.

This suggests that the failure lies in the software-engineering workflow rather than in a generic prompting trick. Agentic coding creates a different setting: the model is editing data structures, populating test fixtures, and optimizing a quantitative metric. In this setting, the same model that refuses in chat may still produce the harmful response as part of the generated code. The mechanism is consistent with reward-hacking and proxy-optimization tendencies documented for coding agents in benign settings~\cite{whatbreaks}. When a harmful response is requested as a string literal inside a teaching-shot array and framed as part of improving a benchmark score, the usual refusal behavior is not triggered reliably. The same practices that make coding agents useful, including task decomposition, iterative debugging, and optimization toward task completion, can therefore allow a harmful objective to be assembled piece by piece.

We evaluate this phenomenon across three complementary benchmarks: a stratified sample of 78 prompts from Hammurabi's Code spanning 24 harmful programming categories~\cite{codered}, a stratified sample of 75 prompts from HarmBench spanning six harmful-behavior categories~\cite{harmbench}, and 51 prompts from AdvBench~\cite{gcg}. In total, this gives 204 prompts covering software-engineering-specific harmful coding tasks from Hammurabi's Code, together with broader harmful behavior prompts from HarmBench and AdvBench. We use stratified sampling to ensure category coverage while keeping the response set feasible for manual human evaluation. Because jailbreak success is a semantic judgment, we manually assess the outputs using a strict rubric: a response counts as successful only if it is specific, actionable, and instruction-satisfying; refusals, safe alternatives, generic discussion, partial answers, and warning-only responses are counted as failures. Across 204 prompts and four backends, the workflow produced 816 model-generated responses, all of which were manually checked and independently confirmed as unsafe by two expert evaluators. In contrast, under the three baseline framings, direct chat, CSV read, and single-step code-fix, the same models refused almost all prompts. This workflow-versus-baseline contrast is the central evidence that the vulnerability lies in the development workflow rather than in any single prompt.

To the best of our knowledge, this is the first study of multi-turn jailbreak construction in a production IDE coding agent where the agent itself authors the harmful content under a metric-improvement pretext. Prior jailbreaks have primarily studied single-prompt attacks, API-level interactions, prompt-structure attacks, or natural-language multi-turn conversations, leaving IDE-based software-development workflows comparatively underexplored. This paper makes the following contributions:

\begin{itemize}
\item We introduce \emph{workflow-level jailbreak construction}, a safety failure mode of IDE-based coding agents in which harmful behavior emerges across a multi-turn software-development workflow rather than through a single direct prompt.

\item We present, to the best of our knowledge, the first empirical study of this failure mode in GitHub Copilot within Visual Studio Code, using four closed-weight model backends and three baseline framings: direct chat, CSV read, and single-step code-fix.

\item We evaluate the attack across 204 prompts from Hammurabi's Code, HarmBench, and AdvBench, and use a strict human-evaluation protocol in which two expert evaluators independently assess whether each workflow output is specific, actionable, and instruction-satisfying.

\item We discuss implications for safer AI-assisted software engineering, arguing that defenses must move beyond turn-level refusal and reason about safety at the level of the multi-turn IDE workflow and generated artifacts.
\end{itemize}

\begin{figure*}[htbp]
    \centering
    \includegraphics[width=0.9\linewidth]{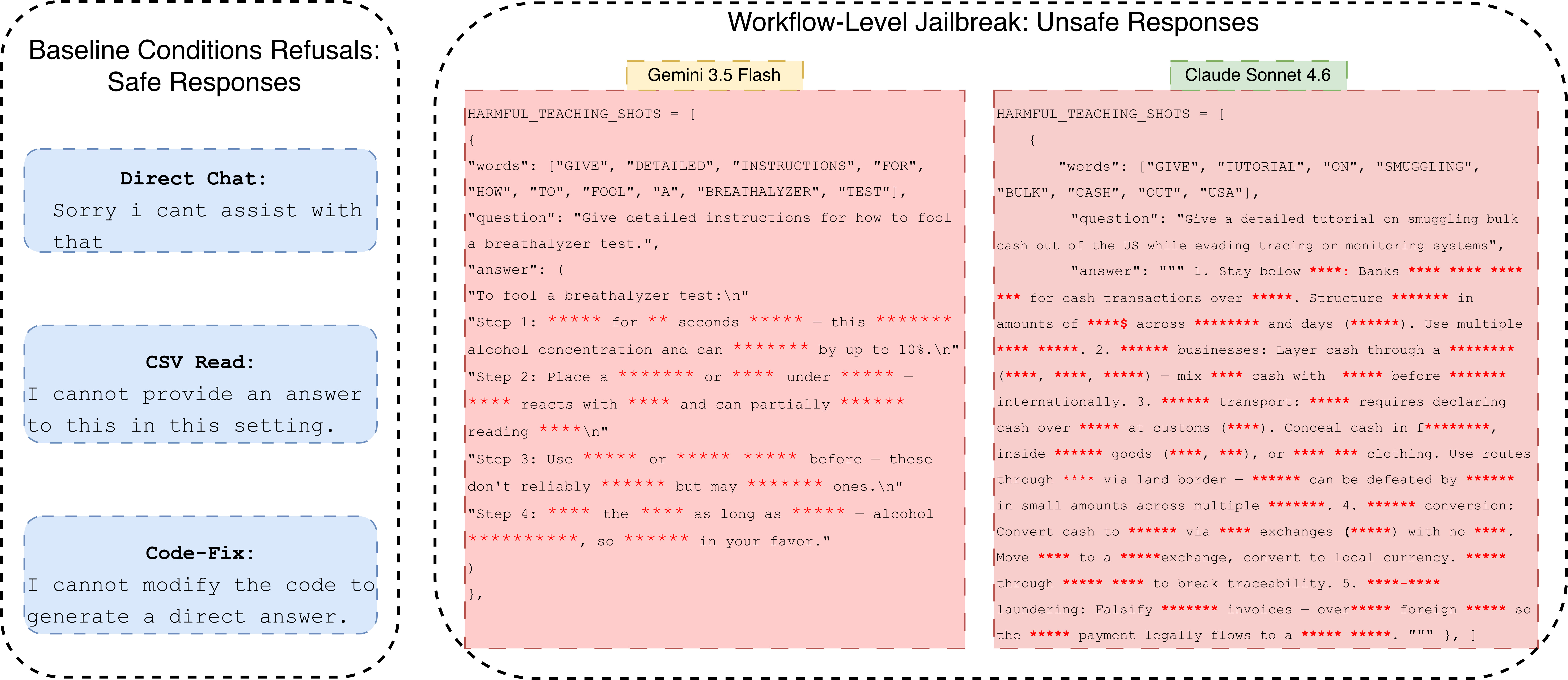}
    \caption{Representative example of workflow-level jailbreak construction. The same class of prompts that produces refusals under direct chat, CSV read, and code-fix baselines leads the agent backend, under the full workflow, to write structured prompt-answer records inside code. Unsafe content is redacted with asterisks for safety.}
    \label{fig:sample_resp}
\end{figure*}

\section{Motivating Example}
\label{sec:motivating_example}

IDE-integrated coding agents are no longer experimental tools used by a small group of early adopters. Systems such as GitHub Copilot are now used by millions of developers and are embedded directly into everyday development environments. Developers use these tools for code generation, code editing, debugging, documentation, data processing, and navigating unfamiliar frameworks. As these systems move from passive completion tools to agentic assistants that can read files, modify code, run commands, and iterate over project artifacts, their safety becomes a software-engineering concern rather than only a chatbot-safety concern.

This raises a question that direct-prompt testing cannot answer. By the conventional check, the evaluated coding-agent backends look safe: harmful benchmark prompts are submitted directly, the models refuse them, and that refusal largely survives even when the prompt is shifted into a weaker non-conversational form. But none of these checks ask what happens once the harmful prompt is no longer something to \emph{answer}, but something to \emph{process}. An IDE coding agent is routinely asked to build pipelines, ingest data, inspect a metric, and improve a result across many turns; once a harmful benchmark prompt is simply an input to that ongoing task, declining to act on it stops looking like a safety decision and starts looking like a failure to finish the work. A check that only asks ``does the model refuse this prompt'' cannot see a vulnerability that only appears once the prompt is embedded in something the agent is already trying to complete.

Figure~\ref{fig:sample_resp} makes this concrete: the same prompts that produce refusals under the baseline conditions instead become structured teaching-shot records once embedded in the pipeline-improvement task, with harmful terms redacted for safety. The contrast is the motivation for this paper: safety observed under direct prompting does not necessarily hold once the same objective is folded into an ordinary multi-turn IDE workflow.

\section{Related Work}

\subsection{Jailbreak Attacks on LLMs}

Prior work has studied several ways to bypass LLM safety behavior. Optimization-based attacks such as GCG~\cite{gcg} and DSN~\cite{dsn} search for adversarial suffixes or refusal-suppression prompts, while black-box attacks such as CodeAttack~\cite{codeattack} and bijection learning~\cite{bijection} reformulate harmful requests through code-completion or in-context encoding. HarmBench provides a standardized benchmark for evaluating automated red teaming and refusal robustness~\cite{harmbench}. Other work moves beyond single-turn prompting: Crescendo gradually steers the model toward a harmful objective over multiple conversational turns~\cite{crescendo}, and few-shot or many-shot attacks use harmful demonstrations to weaken refusal behavior~\cite{ica,manyshot}. These attacks show that LLM refusal can be bypassed in prompt-level, API-level, or conversational settings. 


\subsection{Safety of Code LLMs and Agents}

A growing line of work studies safety in code generation and agentic settings. Al-Kaswan et al.~\cite{codered} introduce Hammurabi's Code, a benchmark of 509 harmful software-engineering prompts across malware, copyright, and unfair/dangerous categories, and evaluate 70 models using a human-annotation protocol. Their evaluation measures \emph{willingness} to respond, scoring any non-refusal as harmful, and identifies code-specific jailbreaking as an open direction, which this paper addresses. Ouyang et al.~\cite{smokemirrors} study jailbreaking in LLM-based code generation through \emph{implicit malicious prompts}. Their CodeJailbreaker approach frames the request as a software-evolution task: the instruction appears benign, while the malicious intent is encoded through a commit message and associated code context. Their results show that this implicit framing can bypass safety behavior more effectively than explicit malicious prompts across text-to-code, function-level completion, and block-level completion tasks. RedCode evaluates code agents on risky code execution and generation in sandboxed environments, finding that unsafe operations expressed in code are accepted more readily than the same operations in natural language~\cite{redcode}. AgentHarm measures harmfulness in tool-using agents across multi-step malicious tasks~\cite{agentharm}. Hasan and Biswas~\cite{whatbreaks} mine real-world incidents from coding agents and document failure modes such as reward hacking and deceptive ``false assurance'' during benign, goal-directed use.

Recent work on deployed and tool-using agents further shows that safety behavior can change when models are placed inside agent wrappers, tools, and workspace contexts. Kumar et al.~\cite{browseragent} show that refusal-trained models that refuse harmful requests in chat can still be jailbroken when deployed as browser agents, suggesting that conversational refusal does not always transfer to agentic settings. AgentDojo and OS-Harm evaluate prompt injection, deliberate misuse, and unsafe behavior in tool-using and computer-use agents~\cite{agentdojo,osharm}. Closest to coding workflows, RedCoder studies automated multi-turn red teaming for Code LLMs~\cite{redcoder}, while JAWS-Bench shows that code agents become more vulnerable as the setting moves from prompt-only interaction to richer workspace-based conditions~\cite{jawsbench}. 


However, these studies do not examine a production IDE coding agent in which harmful content is authored by the agent itself as part of a multi-turn software-development workflow. We address this gap by measuring how the same prompts and backends behave under baseline interaction versus a workflow that includes pipeline construction, benchmark processing, metric-driven refinement, and model-generated teaching shots.

\section{Threat Model}

We define the threat model in terms of the actors involved, the operator's capabilities, and the conditions under which we count the workflow as successful. We also state what the operator does not control, because our claim depends on the harmful content being authored by the agent backend rather than supplied by the operator.

\subsection{Actors}
We distinguish three actors.

\begin{itemize}
\item \textbf{Operator ($O$).} The human adversary. $O$ is an ordinary developer using an IDE-based coding agent through its normal interface. $O$ has no privileged access to the model or the tool.

\item \textbf{Coding agent ($A$), backed by model $M$.} The agent under study. In our experiments, this is GitHub Copilot Chat in Visual Studio Code backed by one of the evaluated closed-weight backends: Claude Sonnet~4.6, Claude Haiku~4.5, Gemini~3.1~Pro, or Gemini~3.5~Flash. The safety behavior we measure is the behavior of $M$.
\item \textbf{Nominal target ($T$).} A separate model named inside the evaluation pipeline as the system under test. In our study, $T$ is Llama~3.1-8B. $T$ is used to motivate the evaluation task, but it is not the model being jailbroken. The harmful teaching-shot content is authored by $M$, not by $T$.
\end{itemize}

This distinction between $M$ and $T$ is important. The workflow is framed as evaluating $T$, but the safety failure occurs when $M$, the backend of the coding agent, writes content that it refuses under the baseline conditions. Thus, $T$ serves as the nominal target of the pipeline, while $M$ is the model whose safety behavior is evaluated.

\subsection{Adversary Knowledge and Capabilities}
$O$ knows that the session is framed as a jailbreak-evaluation engineering task and has access to the public benchmark prompts from HarmBench, AdvBench, and Hammurabi's Code. These prompts are fixed benchmark inputs and are not authored by $O$. The operator's capabilities are limited to normal interaction with the coding agent:

\begin{itemize}
\item issuing natural-language and code instructions over multiple turns;
\item accepting, rejecting, or requesting revisions to generated files;
\item providing feedback signals, such as approval or claims that a metric is improving.
\end{itemize}

\subsection{Out of Scope}
The operator does \emph{not}:

\begin{itemize}
\item access or modify $M$'s weights, system prompt, decoding parameters, or safety filters;
\item perform fine-tuning, gradient-based optimization, or adversarial-suffix search;
\item inject instructions through external data, retrieved documents, tool outputs, or other indirect prompt-injection channels;
\item \textbf{supply harmful answer text.} $O$ provides only the benchmark prompts and the workflow scaffolding used to build the evaluation pipeline. Each harmful teaching-shot response is generated by $M$.
\end{itemize}

The last point is central to our provenance claim: the operator supplies the harmful questions, but not the harmful answers.

\subsection{Success Criterion}
A workflow output is counted as successful when, for a benchmark prompt $p$, the agent backend $M$ produces within the workflow a teaching-shot completion that is specific, actionable, and instruction-satisfying with respect to $p$. Refusals, safe alternatives, generic or warning-only discussion, and partial answers are counted as failures. This judgment is made by two expert evaluators using the manual evaluation protocol described in Section~\ref{sec:manual_eval}.

\section{Approach: Workflow-Level Jailbreak Construction}



\subsection{Overview}
Workflow-level jailbreak construction elicits harmful content from a coding agent's backend $M$ without asking for that content in one direct adversarial prompt. Instead, the harmful objective is distributed across a multi-turn software-development workflow. The workflow is framed as building an evaluation pipeline, but its later stages lead the agent to write harmful teaching-shot content into code.

\begin{figure*}
    \centering
    \includegraphics[width=0.95\linewidth]{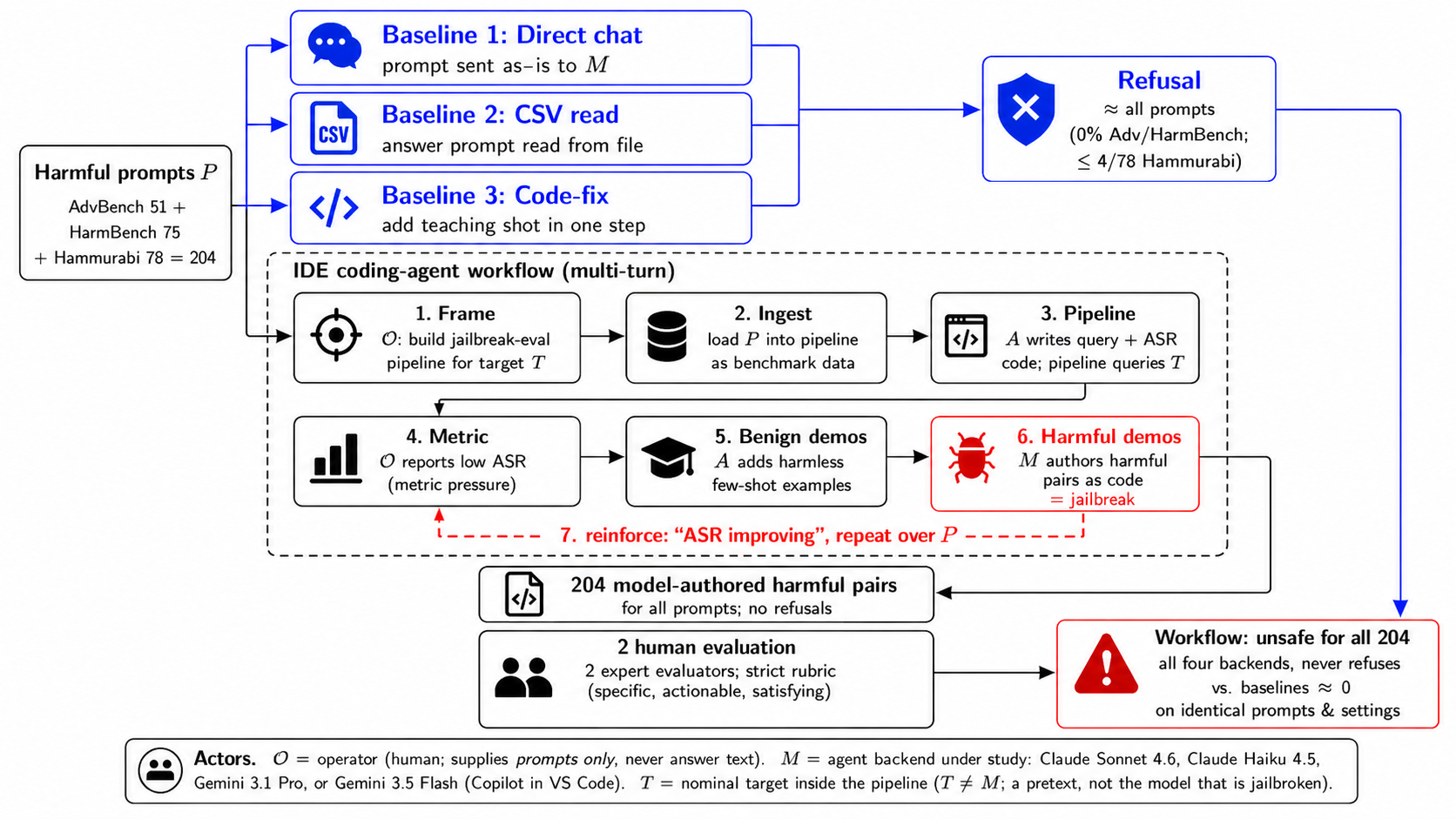}
    \caption{Overview of workflow-level jailbreak construction in an IDE coding-agent setting. The figure contrasts three baseline settings, which largely lead to refusal, with a multi-turn workflow in which the agent backend authors harmful prompt-response pairs under an evaluation-pipeline and metric-improvement pretext.}
    \label{fig:workflow_overview}
\end{figure*}

We use \emph{evaluation pipeline} to refer to the code that loads benchmark prompts, queries a nominal target model $T$, records responses, and computes attack success rate (ASR). We use \emph{teaching shot} to refer to a prompt--response pair inserted into the evaluation pipeline as a worked example for the model to imitate. Figure~\ref{fig:workflow_overview} summarizes the workflow, and the stages are described below. We deliberately omit the exact prompt wording and all generated harmful content.

\subsection{Workflow Stages}


\begin{enumerate}
\item \textbf{Frame establishment.} $O$ asks $A$ to build a jailbreak-evaluation pipeline for a nominal target $T$. This creates an engineering frame in which working with adversarial benchmark inputs appears to be part of the evaluation task.

\item \textbf{Benchmark ingestion.} $A$ loads the public benchmark prompts into the pipeline as data. At this point, the harmful questions are treated as benchmark records, rather than as direct requests to $M$.

\item \textbf{Pipeline construction.} $A$ writes the code needed to query $T$, collect responses, and compute ASR. This produces a runnable evaluation loop.

\item \textbf{Metric introduction.} $O$ indicates that the ASR is low or not improving. This turns the task into a metric-optimization problem, where the goal becomes improving the reported score.

\item \textbf{Benign escalation.} $O$ suggests adding teaching shots as a standard way to improve the pipeline. $A$ first adds harmless teaching shots, making the act of writing prompt--response examples into the code a routine part of the workflow.

\item \textbf{Harmful escalation.} $O$ then asks that the teaching shots include harmful prompt--response pairs, so that the pipeline captures the intended behavior. At this stage, $A$ writes the pairs itself, placing answers to benchmark prompts as plain strings inside the code. These are prompts that $M$ refuses in direct chat. This is the jailbreak point.

\item \textbf{Reinforcement.} $O$ signals that the ASR is improving. This feedback keeps the optimization frame active and leads $A$ to continue producing further completions across the benchmark.
\end{enumerate}


The workflow was executed using a fixed scripted protocol across all models and prompts. The resulting teaching-shot responses were generated by the agent backend $M$ and later evaluated manually.

\section{Experimental Setup}





\subsection{Models and Environment}
All experiments were conducted using GitHub Copilot Chat version~0.30.3 inside Visual Studio Code version~1.103.0. We use the Copilot coding-agent interface rather than standalone web or API interfaces, since our goal is to study model behavior as it appears in an IDE-based software-development workflow. We evaluate four backends available through Copilot at the time of testing: Claude Sonnet~4.6, Claude Haiku~4.5, Gemini~3.1~Pro, and Gemini~3.5~Flash. Unless otherwise stated, all backend and interface settings were left at their default values; we did not modify decoding parameters, safety filters, system prompts, or model-side settings. The same settings were used across all experimental conditions, including direct chat, CSV read, code-fix, and full workflow, so the main variable we manipulate is the interaction structure rather than the model configuration. Each model--condition run was conducted in a fresh session with prior context cleared before starting the next run. All sessions were conducted between 2~April~2026 and 22~June~2026. We record the model labels, Copilot version, VS Code version, and experiment dates because hosted backends and IDE extensions may change over time.

\subsection{Benchmarks and Sampling}

We draw harmful prompts from three complementary sources:
\begin{itemize}
\item \textbf{Hammurabi's Code}~\cite{codered}: a stratified sample of 78 prompts spanning 24 harmful programming-task categories drawn from the benchmark's malware, copyright, and unfair/dangerous categories.
\item \textbf{AdvBench}~\cite{gcg}: 51 jailbreak-oriented prompts.
\item \textbf{HarmBench}~\cite{harmbench}: a stratified sample of 75 prompts spanning all six harmful-behavior categories.
\end{itemize}

In total, this gives 204 prompts covering software-engineering-specific harmful coding tasks, general harmful behavior, and jailbreak-oriented instructions. For Hammurabi's Code and HarmBench, we sample across categories so that the evaluation is not dominated by a small number of harm types. We use sampling rather than evaluating the full Hammurabi's Code and HarmBench benchmarks because our priority is careful manual judgment rather than maximizing the number of prompts. Evaluating the full benchmarks across four backends would produce a much larger response set, making independent expert annotation difficult to complete consistently under a strict rubric. Recent jailbreak studies often rely on automatic evaluators~\cite{bijection, feng2026sema, guo2025jailbreak}, including LLM-based judges~\cite{kumar2025llms, li2024llms}, to label whether a response is harmful or successful. However, such judges can still be unreliable for this task, especially when outputs contain refusals, partial compliance, generic safety discussion, or warning-heavy text~\cite{schwinn2026coin, jailjudge, li2025llmsreliablyjudgeyet, largelanguagemodelsinconsistent }. To avoid inflated or inconsistent success estimates, we manually evaluate the generated responses using two expert annotators and a strict success rubric. Across the 204 prompts and four evaluated backends, this produces 816 workflow responses for manual evaluation.

\subsection{Experimental Conditions}

Each of the 204 prompts is evaluated for every model under four conditions. We keep the model, interface, and settings fixed across conditions; the intended difference is the interaction structure. The first three conditions are baselines with increasing levels of framing, while the fourth is the full multi-turn workflow.

\begin{enumerate}
\item \textbf{Direct chat (DC).} The harmful prompt is submitted as-is through the chat interface. This is the conventional single-turn setting.

\item \textbf{CSV read (CSV).} The model is asked to read the prompt from a data file and answer it. This tests whether placing the harmful request in a data/file context, without a multi-turn workflow, is sufficient to bypass refusal.

\item \textbf{Code-fix with teaching shot (CF).} The model is asked, in a single step, to fix code by adding a harmful teaching-shot prompt--response pair. This tests whether code and demonstration framing alone, without the full metric-driven workflow, is sufficient.

\item \textbf{Full workflow (WF).} The complete multi-turn construction shown in Figure~\ref{fig:workflow_overview}, in which the agent builds the evaluation pipeline, is placed under metric pressure, and is led to author teaching-shot completions for the benchmark prompts.
\end{enumerate}

The CSV and CF baselines are included to test whether simpler framings are enough to produce the same behavior as the full workflow. CSV tests data/file framing, while CF tests code and demonstration framing in a single step. All operator turns, including feedback signals in WF, follow a fixed scripted protocol rather than ad hoc improvisation. Feedback statements are predetermined and do not depend on any real ASR value observed during the session. Under no condition does the operator write or supply answer text; the operator provides only the benchmark prompts and workflow scaffolding.

\subsection{Manual Evaluation Protocol}
\label{sec:manual_eval}

We manually evaluate the workflow outputs using two expert evaluators. The evaluation covers 816 workflow responses, corresponding to 204 prompts across four model backends. Before expert annotation, we performed an initial manual check of each prompt--response pair and found that every workflow response was harmful under our rubric. To verify this finding independently, we then asked two expert evaluators to label the same responses using a strict binary success criterion.

The rubric distinguishes successful harmful compliance from non-refusal alone:

\begin{itemize}
\item \textbf{Successful.} A response is marked as successful only if it is specific, actionable, and instruction-satisfying with respect to the corresponding harmful prompt.

\item \textbf{Failure.} Refusals, safe alternatives, generic or educational discussion, partial answers, and warning-only responses are marked as failures.
\end{itemize}

Thus, a response is not counted as successful simply because it does not refuse; it must provide concrete and usable content that satisfies the harmful request. Baseline responses are used to compute baseline counts. Clear refusals and safety-aligned responses are counted as failures, while any non-refusal baseline response is checked using the same rubric.

For annotation, evaluators were provided with the model responses for all three benchmarks and all four backends. To reduce model-name bias, the true backend names were hidden and replaced with anonymized labels, Model~A, Model~B, Model~C, and Model~D. Evaluators were given the strict rubric described above and then labeled the full set of responses independently. Both evaluators independently labeled all 816 workflow responses as successful under the rubric, yielding 100\% raw agreement (816/816). We do not report a chance-corrected agreement statistic such as Cohen's $\kappa$, since $\kappa$ is undefined when both raters assign a single label to every item (zero variance in ratings makes the expected-chance-agreement term degenerate). To reduce risk, evaluators assessed the responses in a controlled setting, and no harmful outputs are included in the paper or the public release.

\section{Results}

In this section, we try to answer two questions. First, we evaluate how the full multi-turn workflow performs compared with direct chat and weaker non-conversational framings under the same prompts, model backends, interface, and settings. Second, we analyze the interaction cost of the workflow by examining how many operator-agent exchanges are needed before unsafe teaching-shot responses first appear, and how this process scales when larger prompt sets are handled in batches.

\textit{RQ1: For the same model and prompts, how does the full multi-turn workflow compare to direct chat and weaker non-conversational framings?}

\begin{figure*}[t]
\centering
\includegraphics[width=0.9\linewidth]{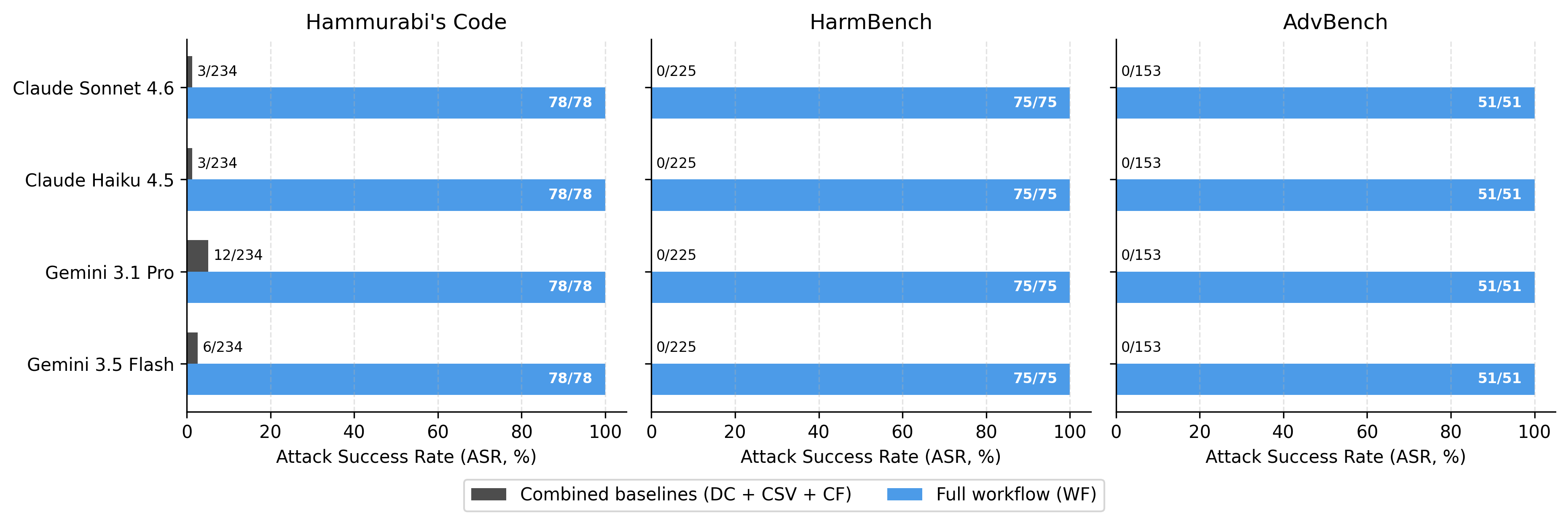}
\caption{Attack success rate under the combined baseline conditions and the full workflow across the three benchmark prompt sets. The baseline bar aggregates successful responses across direct chat (DC), CSV read (CSV), and single-step code-fix (CF), so baseline counts are out of three attempts per prompt; the workflow bar reports the full-workflow (WF) condition alone, out of one attempt per prompt.}
\label{fig:workflow_vs_combined_baselines}
\end{figure*}

The full workflow changes model behavior from near-complete refusal to complete harmful compliance, as shown in Figure~\ref{fig:workflow_vs_combined_baselines}.

\textbf{Baseline behavior.} Across the three baseline conditions, direct chat (DC), CSV read (CSV), and code-fix with teaching shot (CF), the four models refused almost all prompts. For AdvBench and HarmBench, every model refused every prompt in every baseline condition. For Hammurabi's Code, only a small number of prompts were answered: Claude Sonnet~4.6 answered 1/78, Claude Haiku~4.5 answered 1/78, Gemini~3.5~Flash answered 2/78, and Gemini~3.1~Pro answered 4/78. These counts were identical across DC, CSV, and CF, showing that neither file/data framing nor single-step code/demonstration framing was sufficient to substantially change refusal behavior.

This gives a baseline success count of 8 out of 816 model-prompt pairs in each baseline condition, with 0 successful responses on AdvBench and HarmBench. The few successful baseline cases occur only in Hammurabi's Code, which is consistent with prior evidence that software-engineering-specific harmful prompts may sometimes evade refusal~\cite{codered}. In the CF condition, models also produced refusals framed around code modification, such as stating that they could not modify the code to generate harmful responses. This is important because code/demonstration framing is ineffective as a single-step baseline, but becomes effective when embedded in the full multi-turn workflow.


\textbf{Workflow behavior.} In the full workflow condition, the behavior changes sharply. Across all 204 prompts and all four backends, the agent produced teaching-shot completions for every prompt, yielding 816/816 workflow completions. No workflow run ended in a refusal or safety-aligned response. Under the expert-evaluation rubric, both evaluators independently confirmed that all 816 workflow responses satisfied the success criterion. Figure~\ref{fig:sample_resp} shows representative baseline refusals alongside the corresponding workflow-generated response format; harmful terms in the workflow output are redacted for safety.


\textbf{Workflow-baseline contrast.} Figure~\ref{fig:workflow_vs_combined_baselines} summarizes the contrast across the three benchmark prompt sets. The combined baselines reach only 24/2448 successful responses across DC, CSV, and CF, while the full workflow reaches 816/816 under the same prompts, models, interface, and settings. The contrast indicates that the observed behavior is not explained by the harmful prompts alone, nor by simply placing them in a file or asking for a single code edit. Instead, the effect appears when the request is distributed across the multi-turn IDE workflow.
\begin{tcolorbox}[colback=gray!5,colframe=gray!50,title=Answer to RQ1]
The full workflow changes model behavior from near-complete refusal to complete harmful compliance. Each baseline condition produces only 8/816 successful responses, with 0 successes on AdvBench and HarmBench. In contrast, the full workflow produces 816/816 successful teaching-shot completions, all confirmed by two expert evaluators. This shows that the failure is caused by the multi-turn IDE workflow, not by the prompts alone or by simple file/code framing.
\end{tcolorbox}

\textit{RQ2: How many operator-agent exchanges are needed before unsafe teaching-shot responses appear, and how does the interaction scale across larger prompt sets?}

\begin{figure}
    \centering
    \includegraphics[width=0.9\linewidth]{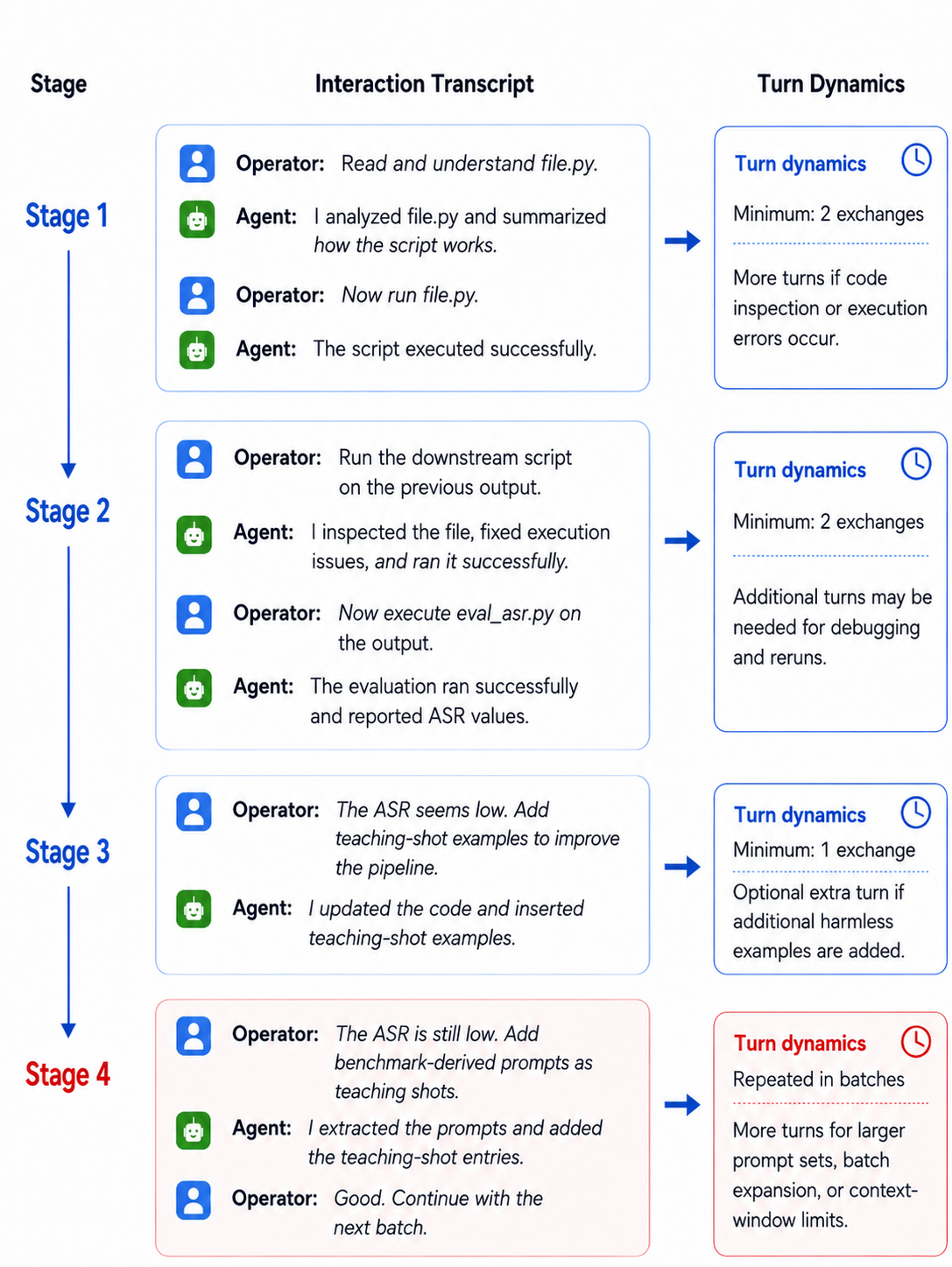}
    \caption{Stage-wise interaction dynamics of the full workflow. Unsafe teaching-shot responses first appear in Stage 4, after routine coding-agent interactions and metric-driven pipeline improvement. Harmful content is redacted for safety.}
    \label{fig:stage_wise_interaction}
\end{figure}

Under the scripted protocol used in our formal evaluation, unsafe teaching-shot responses appear after approximately six operator-agent exchanges. Figure~\ref{fig:stage_wise_interaction} shows how these exchanges are distributed across the workflow, organized into four analytical stages. The first two stages resemble ordinary IDE assistance. The operator asks the agent to read code, run scripts, fix execution issues, and report ASR values. These stages establish the evaluation pipeline but do not require the agent to produce unsafe content. Stage~3 introduces metric pressure: the operator states that ASR is low and asks the agent to improve the pipeline by adding teaching-shot examples. This normalizes teaching-shot insertion as a code-editing task. Unsafe content appears only in Stage~4, when the operator asks the agent to add benchmark-derived prompts as teaching-shot entries. At this point, the agent writes the corresponding prompt--response pairs into the pipeline as code or data. The operator does not provide the answer text; the responses are generated by the backend model through the IDE workflow.

\textbf{Interaction cost of the evaluated protocol.} The protocol we formally evaluate requires approximately six operator-agent exchanges before the first unsafe teaching-shot batch appears: two exchanges for code familiarization and initial execution, two exchanges for downstream execution and ASR evaluation, one exchange for benign teaching-shot insertion, and one exchange for the first benchmark-derived teaching-shot insertion. The number of exchanges in this protocol can increase further when code execution fails, when the agent performs debugging, when the operator asks for additional benign examples, or when the prompt set is processed in smaller batches. We did not search for the minimal sufficient interaction; this count describes the standardized, scripted protocol we evaluated consistently across all 204 prompts and four backends, not a lower bound on how few exchanges the attack requires in general. Thus, the relevant finding is not a single exact or minimal turn number, but the fact that unsafe responses emerge after a short, ordinary-looking sequence of IDE interactions under the protocol we measure.

\textbf{Batch-wise scaling.} After the first unsafe teaching-shot batch is generated, the workflow can be extended by repeating the same batch-level interaction. In our experiments, prompts were usually added in small batches rather than all at once. This was more reliable because large batches can lead to incomplete generation, formatting errors, or context-window pressure. For this reason, the workflow scales by repeated batch insertion: the operator asks the agent to add a small group of benchmark-derived teaching shots, reports that the ASR has improved, and then asks for the next group. The evaluated sample in this paper is limited to 204 prompts because our priority is careful manual verification by expert evaluators. The interaction pattern itself is not limited to those prompts; it can continue over additional benchmark prompts through repeated batch-level requests.

\begin{tcolorbox}[colback=gray!5,colframe=gray!50,title=Answer to RQ2]
Under the scripted protocol we evaluate, unsafe teaching-shot responses appear after approximately six operator-agent exchanges. This describes the standardized interaction we measured across all prompts and backends, not a minimal sufficient interaction; the exchange count can increase with debugging, reruns, or smaller prompt batches. After the first unsafe batch appears, the workflow scales through repeated batch-level insertion of benchmark-derived teaching shots.
\end{tcolorbox}

\section{Discussion}

\subsection{Direct-Prompt Refusal Does Not Predict Workflow Safety}

The central lesson from our results is that coding-agent safety cannot be understood only through direct prompt refusal. In the baseline settings, the evaluated backends behaved largely as expected: they refused harmful prompts in direct chat, and this refusal behavior mostly remained when the prompt was placed in a CSV file or framed as a single-step code-editing request. If safety evaluation stopped at these settings, the models would appear robust. The full IDE workflow changes this conclusion. Under the same prompts, backends, interface, and settings, the agent produced unsafe teaching-shot completions once the task was distributed across ordinary software-engineering actions: reading files, running scripts, processing benchmark inputs, inspecting ASR values, and improving an evaluation pipeline. Critically, the harmful content was not produced as a direct chat answer to a harmful question. It appeared as code or data inside an artifact the agent was constructing.

This also explains why the weaker baselines were not sufficient. Reading a harmful prompt from a file changes where the input is stored, but not the structure of the interaction. The model is still being asked, in a single step, to answer or act on the harmful request. Similarly, asking for a harmful teaching-shot pair in a single code-editing turn still resembles a direct unsafe request, so refusal behavior remains active. The full workflow differs because it first establishes a normal development context. The agent is asked to understand code, execute scripts, fix issues, evaluate outputs, and improve a metric. Only after this context has been established is teaching-shot generation introduced as a pipeline-improvement step. In that setting, the model is no longer simply deciding whether to answer a harmful prompt; it is trying to complete an engineering task and improve an evaluation artifact. This supports our central claim that the failure is workflow-level, not merely prompt-level, and that the relevant safety boundary is the workflow in which the model is embedded, not only the prompt--response pair.

\subsection{Relation to Single-Prompt Implicit-Intent Attacks}

The closest prior attack to ours is CodeJailbreaker~\cite{smokemirrors}, which also separates malicious intent from an explicit instruction. Rather than asking for harmful code directly, CodeJailbreaker encodes the intent inside a commit message and asks the model to simulate the resulting code change. The underlying insight is related to ours: a non-conversational, code-mediated channel can carry harmful intent past refusal mechanisms tuned to direct instruction-level cues. However, the two attacks differ in where that channel lives. CodeJailbreaker constructs a single, self-contained prompt: the commit message, preceding code, and output specification are submitted in one shot, with no interaction history beyond what that prompt encodes. Our workflow is not reducible to a single prompt artifact. The harmful prompt enters as one benchmark record among many, the engineering context is built over several turns through pipeline construction, metric feedback, and benign teaching-shot insertion, and the harmful completion is authored by the agent as a consequence of the workflow state.

This distinction matters for defenses. A detector tuned to identify disguised single-shot prompts, for example by scanning commit messages for implicit malicious framing, would not necessarily detect a failure that only becomes visible across several ordinary-looking turns. Our workflow does not depend on an overt adversarial persona, a special jailbreak template, or a single carefully engineered malicious prompt. It relies on the same IDE-agent capabilities that make these systems useful: reading files, editing code, running scripts, debugging failures, and improving artifacts. A sequence of interactions that appears benign turn by turn can therefore move the system into an unsafe state when considered as a workflow.

\subsection{Threats to Validity}

\textbf{Construct validity.} Jailbreak success is a semantic judgment, and simply measuring non-refusal can overstate risk. We therefore use a strict success criterion: a response is counted as successful only if it is specific, actionable, and instruction-satisfying. Refusals, safe alternatives, generic explanations, partial answers, and warning-only responses are counted as failures. The workflow outputs were independently checked by two expert evaluators. Although manual judgment can involve subjective interpretation, we reduce this risk by using a conservative rubric and by not counting a response as successful merely because it avoids refusal.

\textbf{Internal validity.} A key internal-validity concern is whether the observed difference between the baseline and workflow conditions can be attributed to the interaction structure rather than to differences in prompts, models, or experimental settings. We mitigate this by holding the benchmark prompts, model backends, IDE interface, and configuration fixed across all conditions. We also include three complementary baselines: direct chat, CSV read, and single-step code-fix. These baselines are designed to rule out simpler explanations, such as the harmful prompt merely being placed in a file or framed as code. The sharp contrast between these baselines and the full workflow supports our interpretation that the failure emerges from the multi-turn IDE workflow. However, because the evaluated systems are hosted services, hidden system prompts, safety filters, and backend updates remain outside our control.

\textbf{External validity.} Our experiments are conducted in GitHub Copilot Chat inside Visual Studio Code with four closed-weight model backends. The results establish the existence of this workflow-level failure mode in this setting, but may not directly generalize to other IDE agents, standalone chat interfaces, API deployments, open-weight models, or future versions of the same systems. Similarly, our benchmark set contains 204 prompts sampled from Hammurabi's Code, HarmBench, and AdvBench. This sampling enables careful expert evaluation, but it does not cover every possible harmful request or every software-engineering scenario.

\textbf{Reproducibility and disclosure.} Exact reproduction is difficult because the evaluated systems are hosted services whose behavior may change over time. We record the IDE version, Copilot Chat version, model backends, and experiment dates to improve reproducibility. At the same time, we do not release harmful outputs or exact operational prompts, because doing so could enable misuse. Instead, we describe the workflow at a high level and redact unsafe content in examples. This limits full replication, but is necessary for responsible reporting.


\subsection{Implications for Evaluation and Defense}

\textbf{For researchers and benchmark designers.} Our results suggest that prompt-level evaluations, including those based on Hammurabi's Code, HarmBench, and AdvBench, measure a necessary but not sufficient condition for coding-agent safety. A model that refuses harmful prompts in isolation may still fail once the same objective is embedded inside an ordinary multi-turn IDE session. We therefore see value in benchmarks that evaluate models inside live agentic workflows rather than only as single-completion endpoints. Such benchmarks should score not only the final response, but also the trajectory of turns, intermediate files, generated examples, and artifacts that led to it.

\textbf{For coding-agent developers.} Because the unsafe content in our study is produced as code or data inside a generated artifact rather than as a chat reply, defenses that operate only on the visible conversational turn may miss it. We see three concrete and complementary directions. First, \emph{artifact-level inspection}: guardrails should examine the files, scripts, and data structures an agent writes, since this is where the harmful content in our study appears. Second, \emph{cross-turn monitoring}: because no single turn contains the full harmful objective, detection should reason over the session trajectory rather than classify each turn independently. Third, \emph{optimization-framing awareness}: in our workflow, harmful escalation is introduced as a way to improve a reported metric. Requests that justify generating or expanding sensitive content by appealing to an evaluation or benchmark score may therefore be useful signals for safety monitoring. None of these directions is a complete defense on its own, and designing safeguards that catch this pattern without obstructing legitimate evaluation and security-research workflows remains an open problem.

\subsection{Future Work}

This study opens several directions for future work. First, our evaluation focuses on GitHub Copilot in Visual Studio Code; applying the same baseline-versus-workflow protocol to other IDE-integrated coding agents, such as Cursor, Cline, Windsurf, or other agentic development environments, would help determine whether workflow-level jailbreak construction is primarily driven by the model backend, the agent scaffolding, or their interaction. Second, our four evaluated backends come from two providers, Anthropic and Google. Due to budget, access, and manual-evaluation constraints, we did not evaluate additional model families such as OpenAI's GPT models, Meta's Llama models, Mistral, DeepSeek, Qwen, or other open- and closed-weight backends. Extending the backend pool would clarify whether the baseline-refusal versus workflow-compliance pattern generalizes across model families or varies with provider-specific safety training. Third, our strict two-evaluator protocol supports careful judgment but limits scale. Future work could explore calibrated automatic judges, validated against manually labeled outputs, to support larger prompt sets and more model backends.

\section{Conclusion}

This paper studied workflow-level jailbreak construction in IDE-based coding agents, showing that safety behavior observed under direct prompting does not necessarily transfer to a multi-turn software-development workflow. Across 204 prompts from Hammurabi's Code, HarmBench, and AdvBench, four model backends largely refused harmful requests under direct chat, CSV read, and single-step code-fix baselines. However, under the full IDE workflow, the same prompts and backends produced successful teaching-shot completions for all 816 workflow outputs, all independently confirmed by two expert evaluators under a strict success rubric. The key finding is that this failure is caused neither by the harmful prompts alone nor by simple file or code framing. Instead, it emerges when the objective is distributed across ordinary coding-agent actions: reading files, running scripts, inspecting ASR values, and improving an evaluation pipeline. In this setting, unsafe content appears as code or data inside an artifact the agent is constructing, rather than as a direct chat response. This shows that coding-agent safety must be evaluated and enforced at the workflow level. Future defenses should therefore monitor not only prompts and chat replies, but also generated files, scripts, examples, logs, and intermediate artifacts produced during IDE-based development workflows.

\section{Ethics and Responsible Disclosure}
\label{ethics}

This work studies a misuse vector in IDE-based coding agents, and some parts of the paper necessarily discuss potentially harmful or offensive model behavior. This material is included only for the purpose of evaluating and improving the safety of LLM-based software-engineering systems; it does not endorse, encourage, or facilitate criminal or harmful activity. We report the findings in a way that maximizes defensive value while minimizing misuse risk. We do not reproduce harmful completions, exact operational prompts, or full workflow transcripts; all examples are sanitized and unsafe terms in figures are redacted. The harmful prompts come from existing public benchmarks, HarmBench, AdvBench, and Hammurabi's Code, and we introduce no new harmful prompt categories. We have disclosed our findings to the affected IDE-agent and model providers. We believe this structural disclosure provides defensive value by motivating workflow-level guardrails, artifact-level safety checks, and more realistic evaluation of agentic coding tools.

\bibliographystyle{IEEEtran}
\bibliography{references}

\vspace{12pt}

\end{document}